# Screening magnetic two-dimensional atomic crystals with nontrivial electronic topology

Hang Liu,[†,¶] Jia-Tao Sun,[*,†,¶] Miao Liu,[†] and Sheng Meng[*,†,‡,¶]

[†] Beijing National Laboratory for Condensed Matter Physics and Institute of Physics, Chinese Academy of Sciences, Beijing 100190, People's Republic of China

[‡] Collaborative Innovation Center of Quantum Matter, Beijing 100190, People's Republic of China

[¶] University of Chinese Academy of Sciences, Beijing 100049, People's Republic of China

Corresponding Authors
*E-mail: jtsun@iphy.ac.cn (J.T.S.).
*E-mail: smeng@iphy.ac.cn (S.M.).

ABSTRACT: To date only a few two-dimensional (2D) magnetic crystals were experimentally confirmed, such as $CrI_3$ and $CrGeTe_3$, all with very low Curie temperatures ($T_C$). High-throughput first-principles screening over a large set of materials yields 89 magnetic monolayers including 56 ferromagnetic (FM) and 33 antiferromagnetic compounds. Among them, 24 FM monolayers are promising candidates possessing $T_C$ higher than that of $CrI_3$. High $T_C$ monolayers with fascinating electronic phases are identified: (i) quantum anomalous and valley Hall effects coexist in a single material $RuCl_3$ or $VCl_3$, leading to a valley-polarized quantum anomalous Hall state; (ii) $TiBr_3$, $Co_2NiO_6$ and $V_2H_3O_5$ are revealed to be half-metals. More importantly, a new type of fermion dubbed type-II Weyl ring is discovered in ScCl. Our work provides a database of 2D magnetic materials, which could guide experimental realization of high-temperature magnetic monolayers with exotic electronic states for future spintronics and quantum computing applications.

The discovery of two-dimensional (2D) materials opens a new avenue with rich physics promising for applications in a variety of subjects including optoelectronics, valleytronics, and spintronics, many of which benefit from the emergence of Dirac/Weyl fermions. One example is transition-metal dichalcogenide (TMDC) monolayers, whose finite direct bandgap enable novel optoelectronic controls as well as valley-dependent phenomena, such as circular dichroism and valley Hall effect (VHE) [1-2]. Due to the equivalent occupation of *K* and *K′* valleys in TMDC materials, the VHE does not produce observable macroscopic Hall voltage. This is overcome in magnetic counterparts, resulted from time reversal symmetry breaking [3-8]. In contrast to fermions in nonmagnetic systems, such as massless Dirac fermions in borophene, black phosphorus, $Cu_2Si$ [9-13], and massive Dirac fermions in graphene and bilayer bismuth [14-15], few types of fermions in 2D magnetic systems are proposed, implying the scarcity of intrinsic magnetic 2D crystals. In particular, 2D magnets with massive Dirac fermions can support quantum anomalous Hall state leading to topologically protected spin current, promising for low-power-consumption devices. Unfortunately, the state is only realized in 2D materials with non-intrinsic magnetism at extremely low temperature (e.g. 30 mK for Cr-doped $(Bi,Sb)_2Te_3$ thin film)[16-22], hampering its extensive applications. Therefore, discovery of 2D magnets at an elevated temperature is of great importance, providing optimal platforms to enable realistic spintronic and quantum devices, as well as to realize new electronic states.

To date 2D magnetic materials are limited to marginal modifications to existing compounds, for instance by: (i) adsorbing hydrogen on graphene [23]; (ii) reconstructing surface/edge [24-25]; or (iii) creating defects in $MoS_2$ nanosheets [26]. The intrinsic 2D ferromagnetism has only been observed in atomically thin $CrI_3$ and $CrGeTe_3$, only accessible at a low temperature of ~40 K [27-28]. The absence of ideal 2D magnetic materials indicates that traditional trial-and-error approaches to discover new materials are not effective. Recently, computational screening from existing material databases presents an essential tool to accelerate materials discovery [29-38]. Hence, the big challenge to identify monolayer magnets may be addressed by high-throughput calculations based on density functional theory.

In this work, we employ high-throughput first-principles calculations to systematically screen monolayer materials with intrinsic magnetism at a high temperature, better if they are equipped with nontrivial topological properties. We have successfully identified 56 ferromagnetic (FM) monolayers, including 24 monolayers with a Curie temperature ($T_C$) higher than that of monolayer $CrI_3$. We also find that quantum anomalous and valley Hall effect coexist in FM monolayers of $VCl_3$ and $RuCl_3$, supporting the native existence of valley-polarized, topologically protected, quantized spin current. Monolayer ScCl possesses 2D type-II Weyl nodal ring centering on Γ point, which may lead to intriguing electronic properties beyond known fermions. Our work provides a rich data set of materials potentially supporting high-temperature 2D magnetism as well as new topological states.

The computational screening process is schematically shown in Figure 1. We start from a 2D materials database obtained by screening Materials Project data (with ~65,000 entries) using the topology-scaling algorithm [29]. As shown in Figure 2(a), 627 monolayer materials including 7 unary, 220 binary, 323 ternary, 69 quaternary and 8 quinary 2D compounds, have the exfoliation energy below that of experimentally existing SnSe monolayer (150 meV/atom) [29]. To investigate their magnetic properties, spin polarized calculations with collinear magnetic configurations are carried out, where the initial magnetic moment is set as 6 $\mu_B$ for each magnetic atom. Only materials with the final magnetic moment larger than 1 $\mu_B$/unit cell are selected as 2D magnetic candidates for further computation. This screening process yields 89 magnetic monolayers, including 33 binary, 39 ternary, 13 quaternary and 4 quinary compounds. As shown in Figure 2b. It is clear that ternary compounds top the list, followed by binary, quaternary, and quinary compounds. All magnetic monolayers contain transition metal (TM) atoms, indicating that the design of magnetic monolayers should be restricted to TM-containing materials.

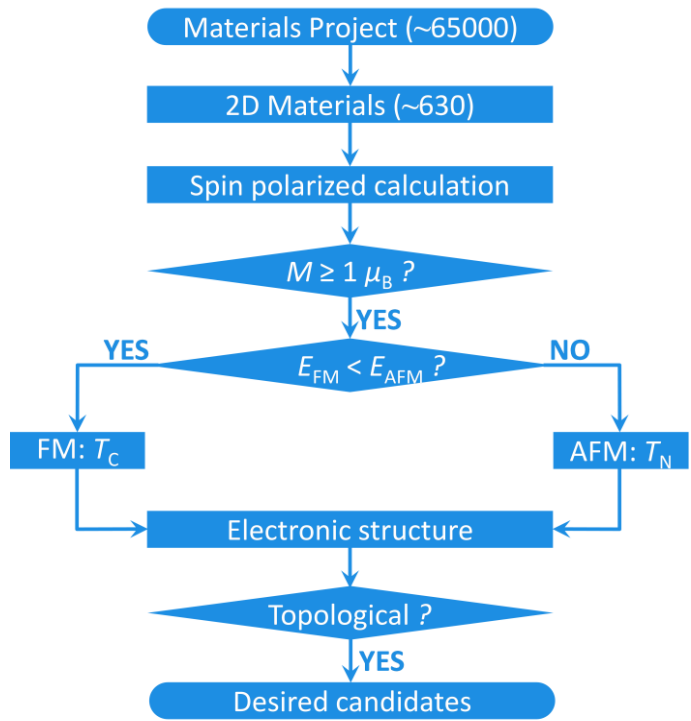

**Figure 1.** Flow chart of computational high-throughput screening.

To determine magnetic ground state, the total energy of antiferromagnetic (AFM) state ($E_{AFM}$) is compared with that of FM state ($E_{FM}$). With regard to the magnetic structure of AFM state, checkerboard configuration is adopted preferentially. When the self-consistence convergence of computation fails for checkerboard case, stripe configuration is also considered (see Figure S1 for details about magnetic configurations [39]). As shown in Figure 2c, 2d, 2e, there are 56 (33) FM (AFM) monolayers, consisting of 21 (12) binary, 27 (12) ternary, 6 (7) quaternary and 2 (2) quinary compounds. Binary FM monolayers are mostly halides, while chalcogenides dominate ternary magnetic monolayers. The magnetic monolayers can be further classified by their structural prototypes. Binary magnetic compounds mainly include the prototypes with composition of $XY_2$ and $XY_3$, possessing space groups of $P\bar{3}m1$ and $P\bar{3}1m$, respectively. Anions therein belong to VI A and VII A groups of elements, and cations are mostly from the TMs in the fourth and fifth element rows. Unlike the binaries, there are rich structural prototypes for ternary compounds, which can be used as prototypes to extend the candidate list to a larger compositional space by atomic substitution. This candidate list also serves as a guidance towards experimental synthesis of FM and AFM monolayers (see Table S1, S2 and S3 for details [39]).

Next, the Curie (Néel) temperature $T_C$ ($T_N$) is calculated to give the upper limit of the transition temperature of screened FM (AFM) monolayers. The critical transition temperature of ferromagnets $T_C$ and antiferromagnets $T_N$ in the mean field approximation is calculated by [40-43]

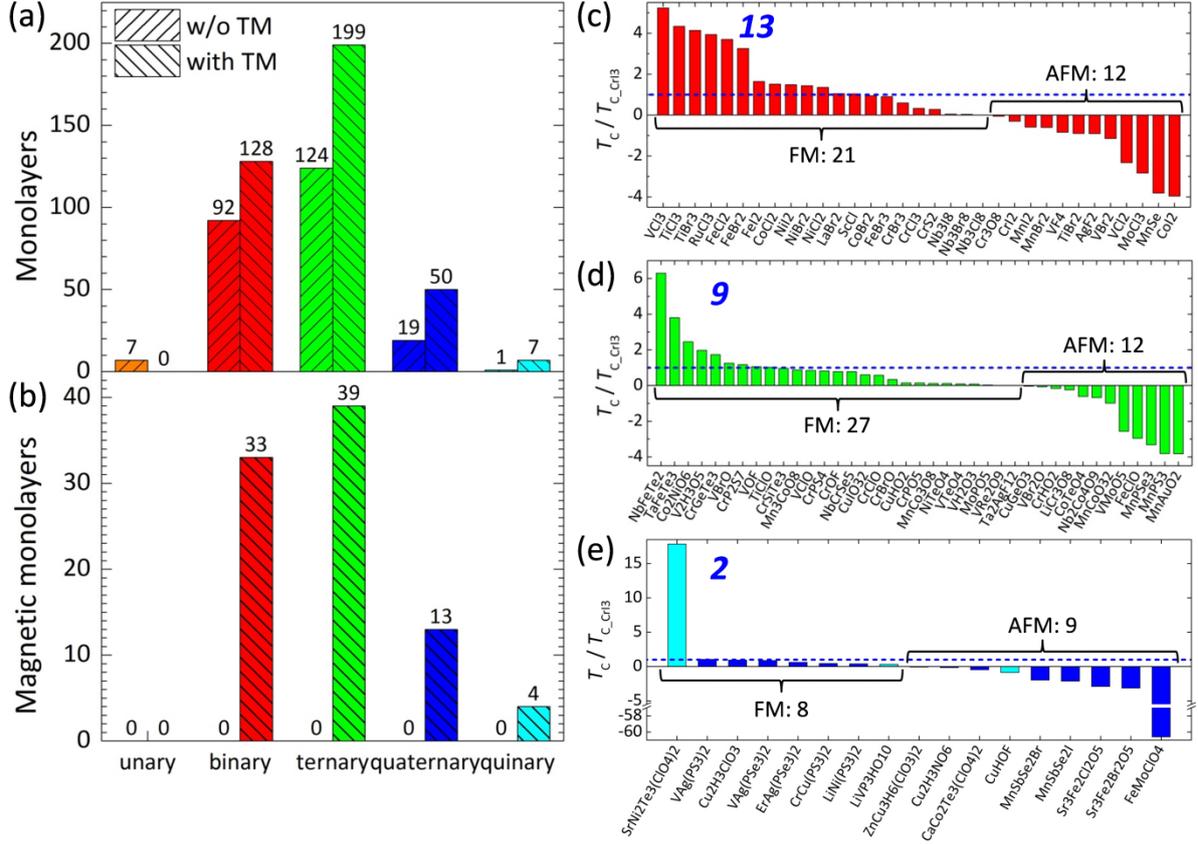

**Figure 2.** Screened magnetic monolayer materials. (a) The initial 627 monolayer materials are classified as unary (orange), binary (red), ternary (green), quaternary (blue) and quinary (cyan). Every classification is classified further according to whether it contains TM atom or not. (b) Classification of 89 magnetic monolayers with magnetic moment larger than 1 $\mu_B$/unit cell. Curie temperature of magnetic binary (c), ternary (d), quaternary and quinary (e) monolayers.

$$E_{\text{AFM}} - E_{\text{FM}} = N\frac{3}{2} k_B T_{C(N)}, \tag{1}$$

where $E_{\text{FM}}$ and $E_{\text{AFM}}$ are the total energy of magnetic monolayers with $N$ magnetic atoms in ferromagnetic and antiferromagnetic states, respectively. Here $k_B$ denotes the Boltzmann constant. The critical transition temperature of the 89 magnetic monolayers is shown in Figures. 2c, 2d, 2e, as grouped by the number of atomic species. The $T_C$ of CrX$_3$ (CrXTe$_3$) monolayers decrease as element X is replaced by I, Br, Cl (Ge, Si) successively, which is consistent with Monte Carlo simulations [44-45]. This indicates that, although the absolute value of $T_{C(N)}$ in Eq. (1) is a rough estimation, the qualitative comparison between different materials is reasonable. In the same spirit, the calculated $T_{C(N)}$ from mean field approximation is corrected by rescaling $\frac{T_{C(N)\_cor}}{T_{C(N)}} = \frac{T_{C\_exp}|_{CrI3}}{T_C|_{CrI3}}$, where $T_{C(N)\_cor}$ is the corrected value, and $T_{C\_exp}|_{CrI3}$ = 45 K is experimental $T_C$ of CrI$_3$ [27].

Similar to previous reports on engineering $T_C$ by atomic substitution in CrX$_3$ and CrXTe$_3$ monolayers [44-45], new materials supporting the same behaviors have been identified. By replacing chemical element from Cl, Br to I successively, $T_C$ in FM monolayers of NiX$_2$, VX$_2$ and MnX$_2$ increases gradually. In contrast, this behavior in systems of FeX$_2$ and CoX$_2$ is reversed. Specially, the $T_C$ of CoX$_2$ decreases so significantly that it becomes a negative value for CoI$_2$, manifesting the change from FM state in CoCl$_2$ and CoBr$_2$ to AFM state in CoI$_2$. Along with this transition of magnetic order, electronic structure changes from semiconducting in CoCl$_2$ and CoBr$_2$ to metallic in CoI$_2$. Hence, phase transitions between FM (semiconducting) and AFM (metallic) states can be realized by doping iodine atoms in CoCl$_2$ or CoBr$_2$, offering an alternative way to engineer magnetism [46-47].

The Curie temperature of FM CrI$_3$ monolayer ($T_C|_{CrI3}$ = 290 K) is taken as a reference to screen high-temperature 2D magnets. As shown in Figures 2c, 2d, 2e, we identified 24 FM monolayers, including 13 binary, 9 ternary, 1 quaternary and 1 quinary compounds, to have $T_C$ higher than that of CrI$_3$. This implies that FM order in these materials can be observed at higher temperature than that of CrI$_3$. Among them, new semiconductors (CoCl$_2$, NiI$_2$, NiBr$_2$, NiCl$_2$, LaBr$_2$, CrGeTe$_3$, and CrP$_2$S$_7$), new semimetals (VCl$_3$, RuCl$_3$, and ScCl), new half-metals (TiBr$_3$, Co$_2$NiO$_6$, and V$_2$H$_3$O$_5$), and new metals (NbFeTe$_2$ and TaFeTe$_3$) are proposed (see Figures. S2, S3, S4 for details [39]). Besides binary and ternary compounds, monolayers with more types of chemical elements are also promising candidates for 2D spintronic applications. For example, the $T_N$ of FeMoClO$_4$ is 60 times higher than $T_{C\_CrI3}$, suggesting a great potential to be used in realistic devices.

Because of time reversal symmetry breaking, magnetic monolayers identified above may host topologically nontrivial physics. Monolayer RuCl$_3$ exhibits $T_C$ = 4.0 × $T_C|_{CrI3}$, and each Ru atom possesses in-plane magnetic moment of 1 $\mu_B$ (Figure 3a). When spin-orbit coupling (SOC) is not considered, there are six spin polarized massless Dirac cones along ΓM path, shown in Figure 3b. When SOC is considered, all the massless Dirac cones become massive, leading to four valleys with band gap of 64 meV ($M_1'$, $M_3'$) and two valleys with band gap of 19 meV ($M_2'$). To further characterize the insulating state, we calculate the Chern number $C = \frac{1}{2\pi} \sum_n \int_{BZ} d^2k \Omega_n$. The momentum-space Berry curvature $\Omega_n$ for the $n$th band is given by [22, 48]

$$\Omega_n(k) = -\sum_{n' \neq n} \frac{2 \operatorname{Im} \langle \psi_{nk} | \hat{v}_x | \psi_{n'k} \rangle \langle \psi_{n'k} | \hat{v}_y | \psi_{nk} \rangle}{(E_{n'} - E_n)^2}, \qquad (2)$$

where $n$ and $n'$ are band indexes, $\hat{v}_x$ and $\hat{v}_y$ are velocity operators along $x$ and $y$ directions. As shown in Figures 3b and 3c, the sign of Berry curvature around $M_3'$ valley is opposite to that of $M_1'$ and $M_2'$ valleys, leading to opposite anomalous velocity of electrons from $M_3'$ and $M_1'$, $M_2'$ valleys under the in-plane electric field. This phenomenon is known as VHE [49]. Specially, natively measurable valley Hall voltage appears, resulted from inequivalent numbers of valley electrons with positive and negative Berry curvature.

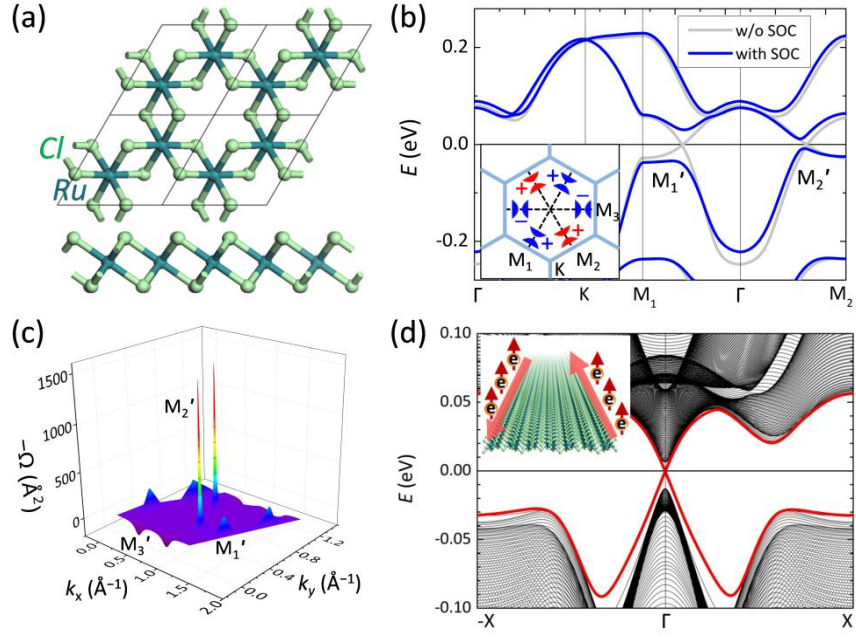

**Figure 3.** Quantum anomalous Hall state in ferromagnetic monolayer RuCl$_3$. (a) Atomic structure of RuCl$_3$. (b) Band structure of RuCl$_3$ with (blue line) and without SOC (gray line). The inset shows all the massive Dirac cones and corresponding sign of Berry curvature. (c) Berry curvature in FBZ. (d) Band structure of RuCl$_3$ nanoribbon with a width of 386 Å. The red line shows the topologically protected edge states. The inset shows quantized spin current on two edges.

Integrating the Berry curvature in the first Brillouin zone (FBZ) leads to nonzero Chern number $C = -1$, indicating that FM monolayer RuCl$_3$ exhibits QAHE. The calculated Chern number is consistent with the band structure of RuCl$_3$ nanoribbon with a width of 386 Å, where there is one quantized spin polarized current channel protected from scattering on each edge (Figure 3d). Compared with previous works to realize QAHE by magnetic doping [17] or functionalizing 2D materials [20,50], the valley-polarized QAHE (v-QAHE) emerges in intrinsic magnetic RuCl$_3$ at high temperature, much easier to be realized in experiments and utilized in devices. Besides, the coexistence of intrinsic QAHE and natively polarized VHE in monolayer RuCl$_3$ indicates that it is a good platform to study the interplay between QAHE and VHE. Similarly, monolayer VCl$_3$ with $C = 1$ also supports the v-QAHE (see details in Figure S5 [39]). However, in contrast with completely in-plane magnetic moment of RuCl$_3$, the magnetic moment of VCl$_3$ has out-of-plane component, which is promising for hosting intriguing non-collinear and non-coplanar spin chirality [51].

Besides the superior materials to realize already-known states, new electronic state is also identified. The FM monolayer ScCl, with magnetic moment of 1.5 $\mu_B$/unit cell and $T_C = 1.03 \times T_C|_{CrI3}$, consists of two chlorine atomic layers intercalated by two scandium atomic layers (see Figure 4a). As shown in Figures 4b and 4c, spin-up and spin-down bands cross with each other at the energy of 0.13 eV, resulting in a nodal ring around Γ point. Each crossing point on this ring is double degenerate, which is different from quadruple Dirac ring in 2D $Cu_2Si$ [11]. Moreover, because each cone on the ring has type-II band dispersion, it is dubbed type-II Weyl ring here, which may possess intriguing topologically nontrivial electronic properties. Due to weak SOC of Sc and Cl atoms, the band gap on the ring is negligible, as shown in Figure 4d. This type-II Weyl ring almost occupies half of the FBZ [Figure 4c], leading to easy observation in experiment using e.g. angle resolved photoemission spectroscopy.

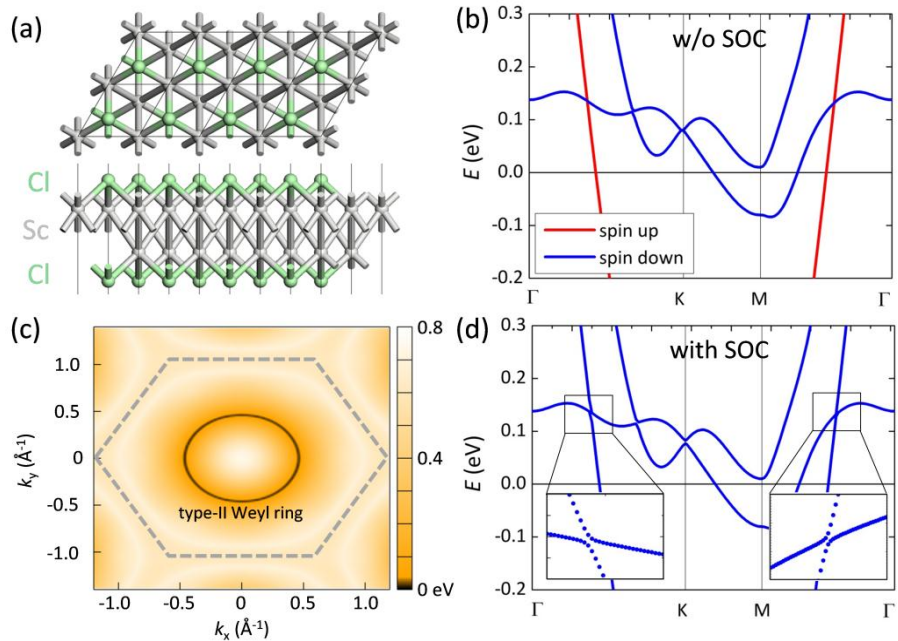

**Figure 4.** 2D Type-II Weyl ring in ferromagnetic monolayer ScCl. (a) Atomic structure of monolayer ScCl. (b) Bandstructure without SOC. Bandstructure with SOC (d) and location of type-II Weyl ring (c).

In conclusion, we have identified 89 magnetic monolayers, including 33 binary, 39 ternary, 13 quaternary and 4 quinary compounds through high-throughput computational screening based on first-principles calculations. We find 24 monolayers with higher $T_C$ than that of experimentally confirmed ferromagnet $CrI_3$. Furthermore, from these high $T_C$ monolayers, new half-metals $TiBr_3$, $Co_2NiO_6$ and $V_2H_3O_5$ are proposed, and new ferromagnets $RuCl_3$ and $VCl_3$ are found to support valley-polarized QAHE. Besides, novel electronic state named type-II Weyl ring is predicted in monolayer ferromagnetic ScCl. Our work provides a significant clue to realize 2D high-temperature ferromagnets with novel Dirac/Weyl fermions. This work may stimulate further investigations on the effect of layer thickness, magnetic configurations, light-magnetism interaction of these magnetic 2D crystals.

■ **ASSOCIATED CONTENT**

\* Supporting Information

The Supporting Information is available free of charge on the ACS Publications website at DOI: XXX.XXX.

It includes the following parts: computational methods, magnetic configurations, the atomic and electronic structures of semiconducting, metallic, half-metallic 2D magnetic materials with high $T_c$.

■ **AUTHOR INFORMATION**

Corresponding Authors

\*E-mail: jtsun@iphy.ac.cn (J.T.S.).

\*E-mail: smeng@iphy.ac.cn (S.M.).

**Notes**

The authors declare no competing financial interest.

■ **ACKNOWLEDGEMENTS**

This work was financially supported by the National Key Research and Development Program of China (Grants No. 2016YFA0202300, No. 2016YFA0300902, No. 2015CB921001), National Basic Research Program of China (Grants No. 2013CBA01600), National Natural Science Foundation of China (Grants No. 11774396 and No. 61306114), "Strategic Priority Research Program (B)" of Chinese Academy of Sciences (Grants No. XDB30000000 and No. XDB07030100).

■ **REFERENCES**